# 18F-FDG brain PET metabolism in post-SARS-CoV-2 infection: substrate for persistent/delayed disorders?

Eric Guedj ( ✉ eric.guedj@ap-hm.fr )
 AMU APHM   https://orcid.org/0000-0003-1912-6132

Matthieu Million
 IHU

Pierre Dudouet
 IHU

Hervé Tissot-Dupont
 IHU

Fabienne Bregeon
 AMU APHM

Serge Cammilleri
 AMU APHM

Didier Raoult
 IHU

Short Report

# Abstract

**Purpose:** Several brain complications of SARS-CoV-2 infection have been reported. It has been moreover speculated that this neurotropism could potentially cause a delayed outbreak of neuropsychiatric and neurodegenerative diseases of neuroinflammatory origin. A propagation mechanism has been proposed across the cribriform plate of the ethmoid bone, from the nose to the olfactory epithelium, and possibly afterwards to other limbic structures, and deeper parts of the brain including the brainstem.

**Methods:** Review of clinical examination, and whole-brain voxel-based analysis of $^{18}$F-FDG PET metabolism in comparison to healthy subjects (p-voxel<0.001, p-cluster<0.05), of two patients with confirmed diagnosis of SARS-CoV-2 pneumonia explored at the post-viral stage of the disease.

**Results:** Hypometabolism of the olfactory/rectus gyrus was found on the two patients, especially one with 4 weeks prolonged anosmia. Additional hypometabolisms were found within bilateral amygdala, hippocampus, cingulate cortex, thalamus, pons and medulla brainstem in the other patient who complained of delayed onset of an atypical painful syndrome.

**Conclusion:** These preliminary findings reinforce the hypotheses of SARS-CoV-2 neurotropism through the olfactory bulb, and the possible extension of this impairment to other limbic structures and to the brainstem. $^{18}$F-FDG PET hypometabolism could constitute a cerebral quantitative biomarker of this involvement. Post-viral cohort studies are required to specify the exact relationship between limbic/brainstem hypometabolisms and the possible persistent disorders, especially involving cognitive or emotion disturbances, residual respiratory symptoms or painful complaints.

# Introduction

Several brain complications of SARS-CoV-2 infection have been already reported including acute cerebrovascular disorders, encephalopathy, encephalitis, and Guillain-Barré syndromes [1]. It has been moreover speculated that SARS-CoV-2 neurotropism could potentially cause a delayed outbreak with onset and progression of neuropsychiatric and neurodegenerative diseases of neuroinflammatory origin [2]. A recent systematic review and meta-analysis of other coronavirus infections confirms confusion, depressed mood, anxiety, impaired memory or insomnia in 27.9 to 41.9% of patients in the acute illness, while in the post-illness stage, the prevalence of post-traumatic stress disorder was of 32.2%, and almost 15% for depression and anxiety disorders [3], with also a clinical overlap with fibromyalgia and chronic fatigue syndrome [4].

As previously shown for other SARS-CoV infections, a propagation mechanism has been proposed across the cribriform plate of the ethmoid bone, from the nose to the olfactory epithelium where ACE2 receptors are highly expressed [5]. This viral neurotropism through the olfactory bulb could be especially responsible of the anosmia frequently reported in these patients [5]. Accordingly, a cortical FLAIR-MRI hyperintensity was visually identified in the right gyrus rectus and olfactory bulbs of a patient with SARS-CoV-2 anosmia [6]. By trans- synaptic transfer, again already reported for other virus [7], this propagation from the olfactory bulb could spread to other limbic structures, such as the amygdala, the hippocampus, and the cingulate cortex which are well-known to be involved in cognition and emotion [8], and pathologically in neurodegenerative and neuropsychiatric diseases [2]. SARS-CoV-2 may also target deeper parts of the brain including the thalamus and the brainstem, and then potentially contributes to the respiratory impairment [7], and also to painful symptoms.

We report here the clinical case of two patients with confirmed diagnosis of SARS-CoV-2 pneumonia, which were explored by whole-body 18F-FDG PET in the post-illness stage to metabolically characterize residual lung abnormalities. Brain PET metabolism was statistically analyzed in comparison to healthy subjects at whole-brain voxel-based level (proportional scaling), and findings were confronted to the hypothesis of the propagation of SARS-Co-2 from the olfactory bulb to other limbic structures, and possibly the brainstem.

## Case Reports

We first report the case of a 54-year-old man with a SARS-CoV-2 pneumonia confirmed by nasopharyngeal swab. We noticed type 2 diabetes, hypertension, and minor asthma in his past medical history. The pneumonia was complicated with a severe acute respiratory distress syndrome requiring 6 days of mechanical ventilation and 15 days of hospitalization in intensive care unit. The evolution was clinically favorable, with nevertheless the persistence of anosmia and ageusia during 4 additional weeks, and delayed occurrence of memory complaints 8 weeks after the first pulmonary symptoms. A whole-body 18F-FDG PET was afterwards performed, while the patients only complained of memory impairment, to evaluate the metabolic activity of persistent pulmonary CT condensations. No suspect lung hypermetabolism was found. The brain 18F-FDG PET exhibited hypometabolism of bilateral rectal gyrus prevailing on the right-side, without evident abnormality on the attenuation correction CT scan. A whole-brain voxel-based SPM8 comparison, to a local database of 23 healthy subjects with same median age as the patient (±10 years), confirmed this hypometabolism including the right olfactory bulb, with interestingly no remote abnormalities to point towards another brain gateway (Figure 1; $p\text{-voxel}<0.001$, k=286, p- cluster=0.023, T-score_max=5.97; BA11/47; peak MNI coordinates: 4 32 -30).

We secondly report the case of a 62-year-old man with a SARS-CoV-2 pneumonia confirmed by nasopharyngeal swab. The patient had not significant antecedent. The pneumonia was complicated with moderate acute respiratory distress syndrome requiring 24 hours hospitalization in intensive care unit and treatment by low-flow nasal oxygen. The patient never experienced anosmia or ageusia. The evolution was clinically favorable with nevertheless the occurrence, 7 days after the first pulmonary symptoms, of a bilateral sensitive lower leg crushing sensation in toes, without any motor or sensory deficiency on clinical examination. A whole-body 18F-FDG PET was afterwards performed, while the patients only complained of this painful syndrome, to evaluate the metabolic activity of persistent pulmonary CT condensations. No suspect lung hypermetabolism was found. The brain 18F- FDG PET exhibited extended bilateral marked hypometabolism especially involving olfactory/rectal gyrus, other limbic structures such as amygdala, hippocampus and cingulate cortex, as well as thalamus, pons and medulla brainstem. These hypometabolisms were confirmed on 4 clusters by a whole-brain voxel-based SPM8 comparison to a local database of 24 healthy subjects with same median age as the patient (±10 years) (Figure 2; p- voxel<0.001, p-cluster < 0.05 corrected). No abnormality was found on the attenuation correction CT scan.

## Discussion

These preliminary findings reinforce the hypotheses of SARS-CoV-2 neurotropism through the olfactory bulb in patients with or without anosmia, and the possible extension of this impairment to other limbic structures and to the brainstem, with brain abnormalities persisting after the remission of the infectious disease. 18F-FDG PET hypometabolism could constitute a cerebral quantitative biomarker of this involvement [9]. Post-viral cohort studies are required to specify the exact relationship between limbic/brainstem hypometabolisms and the possible persistent disorders, especially involving cognitive or emotion disturbances, residual respiratory symptoms or painful complaints. These metabolic features could be also correlated to other brain PET biomarkers of neuroinflammation and infection, such as TSPO and CXCR4 expressions [10], to address the possible relationship with the hypothesized association to neuropsychiatric and neurodegenerative diseases of neuroinflammatory origin [2].

## Declarations

Funding

The local PET database of healthy controls was funded by APHM (regional PHRC; NCT00484523)

Conflicts of interest/Competing interests

The authors declare that they have no conflict of interest

Ethics approval

The two case reports are retrospective observation with no ethical approval requirement other than the informed written consent.

The local PET database of healthy controls was acquired in accordance with the Declaration of Helsinki, with informed written consent of patients and approvement of local ethics committee.

Consent to participate

Informed written consent was obtained from all individual participants included in the study

Consent for publication

Informed written consent was obtained from all individual participants included in the study

Availability of data and material

The PET data that support the findings are available from the corresponding author upon reasonable request

Code availability Not applicable

# Figures

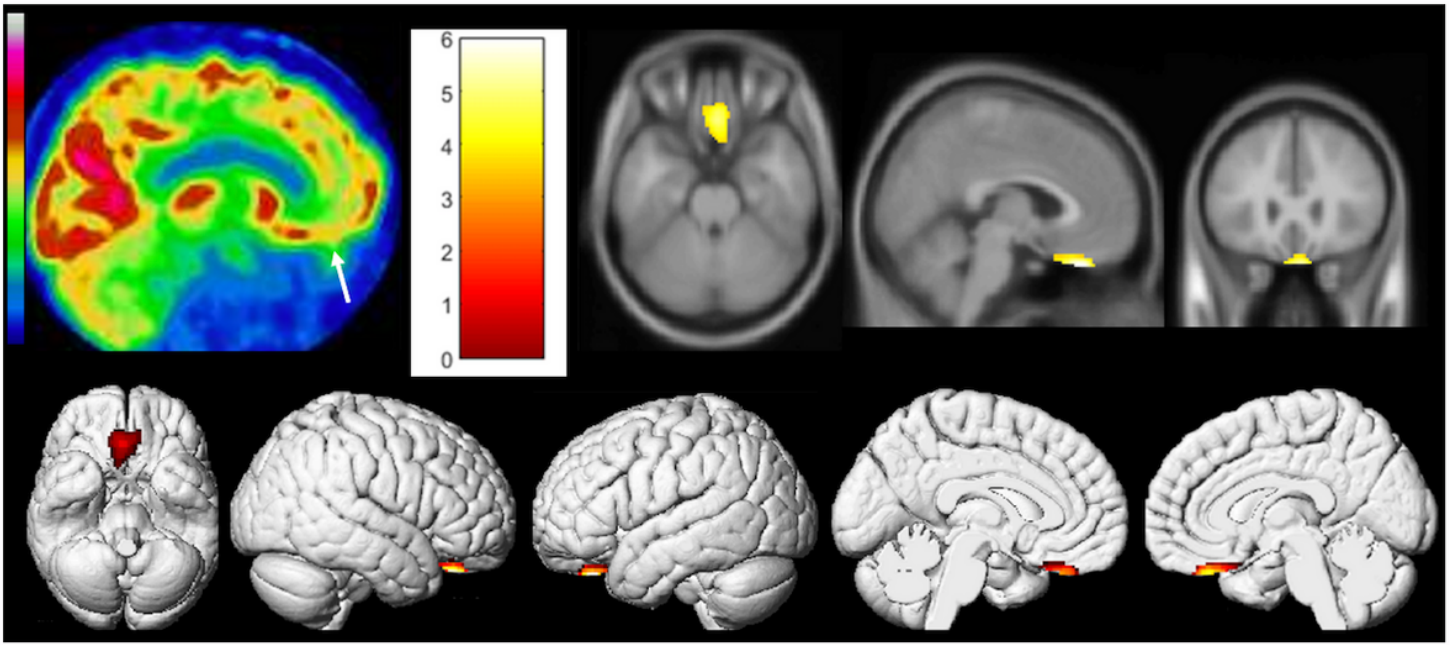

#### Figure 1

Brain 18F-FDG PET hypometabolism of the first patient. Bilateral hypometabolism of olfactory/rectal gyrus is visually identified (white arrow), and confirmed by whole-brain voxel-based SPM8 comparison, to healthy subjects (p-voxel<0.001, p-cluster=0.023).

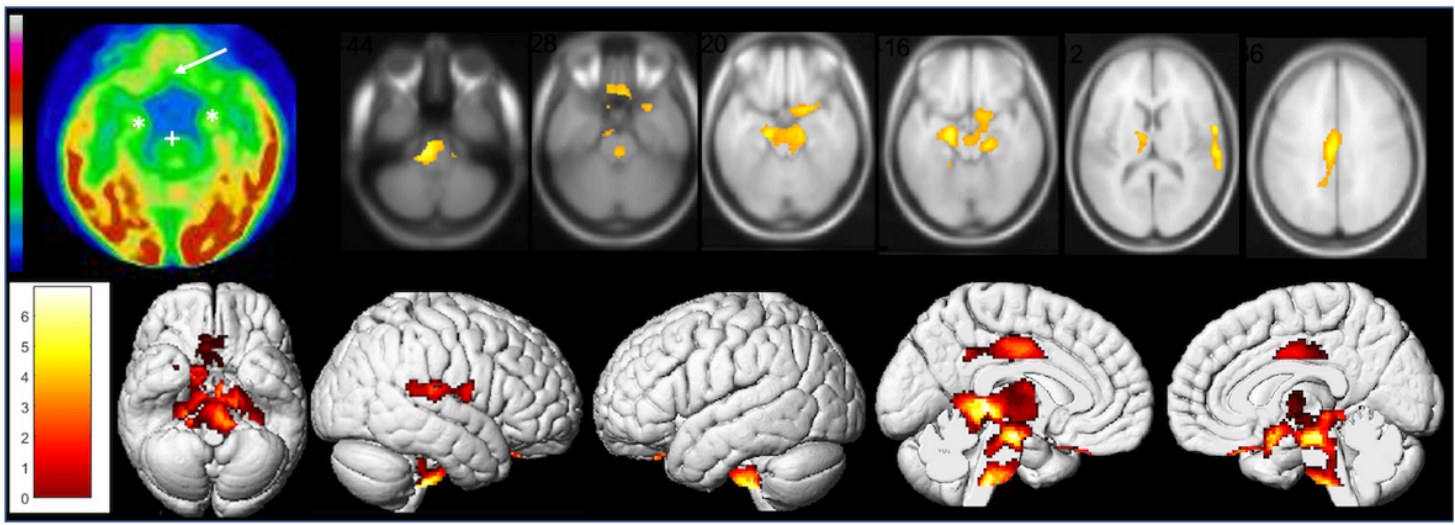

#### Figure 2

Brain 18F-FDG PET hypometabolism of the second patient. Bilateral hypometabolism of olfactory/rectal gyrus (white arrow), medial temporal lobe (white*) and brainstem (white+) is visually identified, and confirmed by whole-brain voxel-based SPM8 comparison to healthy subjects (p-voxel<0.001, p-cluster<0.05).